\documentclass[aps,prd,reprint,superscriptaddress,amsmath,amssymb,amsfonts]{revtex4-1}

\usepackage[boldvectors]{physymb}

\begin{document}

\title{Fermions in Randall-Sundrum models with two additional \\ unwarped extra dimensions}

\author{Erin \surname{De Pree}}
	\email{ekdepree@smcm.edu}
	\affiliation{Department of Physics, St.~Mary's College of Maryland, St.~Mary's City, Maryland, USA}

\author{Jeremy Perrin}
	\email{jdp279@cornell.edu}
	\affiliation{Department of Physics, St.~Mary's College of Maryland, St.~Mary's City, Maryland, USA}
	\affiliation{Department of Physics, Cornell University, Ithaca, New York, USA}

\date{\today}

\begin{abstract}
	
Models with one warped and two unwarped extra dimensions allow for the solution of a number of open questions in particle physics. They can be used to solve the hierarchy problem in the same sense as Randall-Sundrum extra dimensions, they incorporate the Randall-Sundrum approach to flavor, and they generate a dark matter candidate via Kaluza-Klein parity in the flat extra dimensions. In this paper, we examine the models $\textrm{AdS}_{5}\times T_{2}$ and $\textrm{AdS}_{5}\times S_{2}$, deriving the Kaluza-Klein spectrum for fermions propagating in the bulk. While the toroidal model allows for a chiral zero mode, we find that the positive curvature of the spherical model disallows all zero modes without further modifications.

\end{abstract}



\maketitle

\section{Introduction \label{sec:intro}}

	Extra-dimensional models can be used to solve a number of problems in particle physics. In particular, Randall-Sundrum (warped) extra dimensions allow for a natural generation of the Planck-weak \cite{RS} and fermion mass \cite{FermionMass} hierarchies, and orbifolded universal extra dimensions (UED) have a discrete symmetry, known as Kaluza-Klein (KK) parity, that could produce a dark matter candidate  \cite{KKParity}. It is natural to look for ways to combine these positive characteristics into a single model. Gluing together multiple warped throats in a single extra dimension can create a UED-like symmetry that generates a dark matter candidate \cite{Agashe}, but it is also possible to form a product space of warped and flat extra dimensions in a way that combines the characteristics of Randall-Sundrum and UED models.  We examine the behavior of fermions in a few realizations of this latter scenario.
	
\section{Warped Extra Dimension \label{sec:rs}}

	The Randall-Sundrum (RS1) model of warped extra dimensions can be used to generate a natural hierarchy of scales between fixed points in the extra dimension. If the theory has the metric
	\begin{equation}
		\textrm{d}s^{2}=e^{2 k r_{c}|\phi|}\eta_{\mu \nu} \textrm{d}x^{\mu} \textrm{d}x^{\nu}-r_{c}^{2}\textrm{d}\phi^{2}
		\label{eq:RSmetric}
	\end{equation}
(with $\eta_{\mu \nu}=\textrm{diag}(1,-1,-1,-1)$ and $\phi \in (-\pi, \pi) $), and the Higgs boson is confined to the TeV brane, we find that its effective mass parameter is naturally reduced by a factor of $e^{-kr_{c}\pi}$, generating a weak-scale mass from a fundamental mass of order $10^{19}$ GeV \cite{RS}.
	
	In the models we examine in this paper, all of the Standard Model particles except for the Higgs boson are allowed to propagate in the extra dimension(s). For the 5-dimensional case, this allows for the generation of a large hierarchy in fermion masses from a small change in fundamental parameters \cite{GherghettaPomarol,FermionMass}. Without some additional symmetry breaking, however, we run into problems when we try to obtain a chiral 4D fermion. In contrast to the 4D case, a Dirac spinor in 5D is vectorlike, meaning that the `chirality' matrix $\gamma^{5}=i\gamma^{0}\gamma^{1}\gamma^{2}\gamma^{3}$ does not project via $(1\pm \gamma^{5})/2$ onto inequivalent two-component representations of the 5D Clifford algebra (of which there are none). We can still carry out these projections, with the ultimate aim of reinterpreting them as 4D Weyl spinors. However, there is nothing to distinguish these two-component spinors in the fermion action, making it impossible to have a 4D chiral theory.

This problem can be solved by orbifolding the extra dimension. If we require the action to be invariant under the $\mathbb{Z}_{2}$ parity transformation $\phi \rightarrow -\phi$, it turns out that the two `Weyl' spinors must have opposite $\mathbb{Z}_{2}$ parities. Therefore, only one can have a zero-mode (the one with even parity) and the chirality of the 4D theory is restored.

In RS1, the Kaluza-Klein decomposition relates the five dimensional action to a sum over four-dimensional particle actions with varying mass. The 5D action for fermions can be written as
	\begin{align}
		S_{5D} &= \int \textrm{d}^{4}x \int_{-\pi}^{\pi} \textrm{d}\phi \sqrt{G} \left[ \frac{i}{2}\bar{\Psi} E^{A}_{a}\Gamma^{a}\partial_{A}\Psi 
			 \right. \nonumber \\
			 & \qquad \left. 
			-\frac{i}{2}(\partial_{A}\bar{\Psi})E^{A}_{a}\Gamma^{a}\Psi-M\epsilon(\phi)\bar{\Psi}\Psi\right]
		\label{eq:RS1action}
	\end{align}
	In this formula, $G$ is the determinant of the metric, $E^{A}_{a}$ is the inverse vielbein, and $\Gamma^{a}$ is the matrix representation of the 5D Clifford algebra. The $M\epsilon(\phi)$ term is a $\mathbb{Z}_{2}$-odd mass, which we require to be odd to keep the Lagrangian density even under parity transformations.
	
	\subsection{Fermion Decomposition \label{subsec:RS1fermions}}
	
	The 5D fermion kinetic term in equation \ref{eq:RS1action} is split up into two parts so the 5D Dirac operator is Hermitian. In flat, non-compactified spaces, this is identical to the standard form of the fermion action, but here we are left with additional terms in the action that arise from the integration by parts.
	
	Now, we compare the 5D fermion action to the sum over 4D fermions
	\begin{equation}
		\int \textrm{d}^{4}x\sum_{n}\bar{\psi}_{n}\left[i\partial_{\mu}\gamma^{\mu}-m_{n}\right]\psi_{n}
		\label{eq:FermionAction}
	\end{equation}
	We will not go through the KK decomposition procedure explicitly here, since we detail a similar (but significantly more complicated) procedure later in section \ref{sec:6d7d}.  For specifics, see ref.~\cite{GrossmanNeubert, GherghettaPomarol}.  After integrating by parts and decomposing the fermion into a sum over left and right 4D-chiral KK modes,
	\begin{equation}
		\Psi=\sum_{n}\left[\psi^{n}_{R}(x^{\mu})f^{n}_{R}(\phi)+\psi^{n}_{L}(x^{\mu})f^{n}_{L}(\phi)\right]
		\label{eq:FermionDecomposition}
	\end{equation}
	we find by examining the 5D action that $f^{n}_{L}$ and $f^{n}_{R}$ must have opposite $\mathbb{Z}_{2}$ parity. Comparing the 4D kinetic terms of equations \ref{eq:RS1action} and \ref{eq:FermionAction}, we find the following orthonormality conditions:
	\begin{align*}
		\int_{-\pi}^{\pi} r_{c} e^{-3 \sigma}\bar{f}^{m}_{L} f^{n}_{L}\textrm{d}\phi &=\delta_{mn}\\
		\int_{-\pi}^{\pi} r_{c} e^{-3 \sigma}\bar{f}^{m}_{R} f^{n}_{R}\textrm{d}\phi &=\delta_{mn}
	\end{align*}
	Comparing the remaining terms, we find the following pair of coupled differential equations:
	\begin{align}
		\frac{e^{-\sigma}}{r_{c}}\partial_{\phi}\hat{f}^{n}_{R}+e^{-\sigma}M\epsilon(\phi)\hat{f}^{n}_{R}=m_{n}\hat{f}^{n}_{L} 
			\label{eq:RS1diffeq1}\\
		\frac{e^{-\sigma}}{r_{c}}\partial_{\phi}\hat{f}^{n}_{L}-e^{-\sigma}M\epsilon(\phi)\hat{f}^{n}_{L}=-m_{n}\hat{f}^{n}_{R}
			\label{eq:RS1diffeq2}
	\end{align}
	where $\hat{f}=e^{-2\sigma}f$. The general solution to these differential equations can be expressed in terms of Bessel functions of the first kind. 
	For $\phi>0$,
	\[
		\hat{f}^{n}_{L} = e^{\sigma/2} \left[ A J_{\frac{1}{2} - \frac{M}{k}} \left( \frac{e^{\sigma}m_{n}}{k}\right) 
		+ B J_{-\frac{1}{2} + \frac{M}{k}} \left( \frac{e^{\sigma}m_{n}}{k} \right) \right]
	\]
	\[
		\hat{f}^{n}_{R} = e^{\sigma/2}\left[C J_{-\frac{1}{2}-\frac{M}{k}}\left(\frac{e^{\sigma}m_{n}}{k}\right)
		+D J_{\frac{1}{2}+\frac{M}{k}}\left(\frac{e^{\sigma}m_{n}}{k}\right)\right]
	\]
	These two functions must have opposite $\mathbb{Z}_{2}$ parity, and picking a parity choice allows us to proceed. The eigenvalues $m_{n}$ are determined by applying homogeneous Dirichlet boundary conditions to the odd-parity solution. We apply Neumann boundary conditions to the even-parity solution. Because of the $\epsilon(\phi)$ term in the differential equations, these boundary conditions must be non-homogeneous. The constants $A$, $B$, $C$, and $D$ are determined by a combination of boundary conditions and normalization.

However, the solution to equations \ref{eq:RS1diffeq1} and \ref{eq:RS1diffeq2} is much simpler in the $m_{n}=0$ case. We see that they uncouple, and their solutions are (real) exponentials, $f^{n}_{L}=e^{(2k+M)r_{c}|\phi|}$ and $f^{n}_{R}=e^{(2k-M)r_{c}|\phi|}$. Both of these functions are even, so only one of the two can be a physical solution. Assume here for the sake of simplicity that this mode is right-handed. We see that the profile of this zero mode depends strongly on the magnitude of the 5D mass, $M$. If the Higgs boson is localized at the TeV brane, we can control the value of the fermion-Higgs overlap, proportional to $e^{(2k-M)r_{c}\pi}$, by varying the value of $M$. It follows that the 4D fermion mass can be adjusted over a wide range of values while varying the fundamental mass $M$ over less than an order of magnitude, leading to a much more comfortable hierarchy in fundamental fermion masses.

	\subsection{RS1 Limitations \label{subsec:RS1limits}}
	
	We see that Randall-Sundrum extra dimensions provide reasonable solutions to both the Planck-weak and fermion mass hierarchy problems. As it stands, though, this model predicts no stable, weakly-interacting particles that could make up the universe's dark matter. However, we can generate dark matter candidates with different extra-dimensional models. A model known as Universal Extra Dimensions (UED), which is essentially the same as RS1 without the warp factor, has a symmetry that leads to stable Kaluza-Klein modes \cite{KKParity}. Since KK modes have discrete, well-determined values of momentum along the extra dimension, momentum conservation causes KK number to be conserved. Note that, since RS1 does not have translation invariance along the extra dimension, it cannot have momentum conservation.
	
	Even in UED, though, orbifolding the extra dimension breaks the translation invariance and therefore the conservation of KK number. However, we are left with a symmetry known as KK parity. This symmetry forbids the decay of particles of even KK number into particles of odd KK number, and vice versa. Therefore, the lightest particle with a KK number of 1 is both heavy and stable! If this particle is weakly interacting, it could provide an excellent dark matter candidate.
	
	However, Universal Extra Dimensions fails to provide a solution to the Planck-weak hierarchy problem. Therefore, we are motivated to look for models that combine the positive characteristics of UED and RS1. One possible way to do so is by forming a product space of warped and non-warped extra dimensions.

\section{Expanding Beyond 5 dimensions \label{sec:6d7d}}

A six-dimensional model with one warped and one flat extra dimension seems to be the obvious choice, but models of this type have one significant problem: it is impossible to write a bulk mass term for a minimal 6D fermion, which is 6D-chiral and has four components. This is analogous to the case of a two-component Weyl fermion in 4D. Since a bulk mass term cannot be written, we cannot generate a fermion mass hierarchy as we did in RS1. (These types of models are examined in \cite{DavoudiaslRizzo}). 

It may be possible to start with a non-minimal 6D fermion with a bulk mass term. Another possible solution, which we consider here, is to begin with a minimal fermion in a seven-dimensional model.  It turns out, as is demonstrated in \cite{McDonald1}, that we can generate a chiral zero-mode from a seven-dimensional fermion.
	
	At this point, we are faced with another choice. We can consider models with the warp factor acting on all the unwarped coordinates, like
	\[
		\textrm{d}s^{2}=e^{-2kr_{c}|\phi|}\left[\eta_{\mu \nu}\textrm{d}x^{\mu}\textrm{d}x^{\nu}-R^{2}(\textrm{d}\theta_{1}^{2}\\
		+\textrm{d}\theta_{2}^{2})\right]-r_{c}^{2}\textrm{d}\phi^{2}
	\]
We denote this model, which is discussed in \cite{AdS7} and \cite{McDonald2}, by $\textrm{AdS}_{7}$. Alternatively, we can only have the warp factor act on the four ordinary spacetime dimensions:
	\begin{equation}
		\textrm{d}s^{2}=e^{-2kr_{c}|\phi|}\eta_{\mu \nu}\textrm{d}x^{\mu}\textrm{d}x^{\nu}-r_{c}^{2}\textrm{d}\phi^{2}-R^{2}(\textrm{d}\theta_{1}^{2}
		+\textrm{d}\theta_{2}^{2})
		\label{Torus Metric}
	\end{equation}
We denote this model, which is discussed in \cite{McDonald1}, by $\textrm{AdS}_{5}\times T_{2}$. It turns out that the first of these two models, $\textrm{AdS}_{7}$, runs into problems when both fermions and bosons propagate in the bulk, since the couplings between their zero modes are volume-supressed \cite{McDonald2}. Therefore, we consider $\textrm{AdS}_{5}\times T_{2}$ in this paper. In addition, we will examine a variant of this model, with the torus replaced by a sphere:
	\begin{equation}
		\textrm{d}s^{2}=e^{-2kr_{c}|\phi|}\eta_{\mu \nu}\textrm{d}x^{\mu}\textrm{d}x^{\nu}-r_{c}^{2}\textrm{d}\phi^{2}-R^{2}(\textrm{d}\theta^{2}
		+\sin^{2}\theta\textrm{d}\omega^{2})
		\label{Sphere Metric}
	\end{equation}
We refer to this model as $\textrm{AdS}_{5}\times S_{2}$. 

To generate a chiral zero mode, we orbifold both the torus and the sphere by a single discrete symmetry. For the torus, this symmetry takes the point $(\theta_{1},\theta_{2})$ to $(-\theta_{1},-\theta_{2})$, and for the sphere, it takes the point $(\theta, \omega)$ to $(\pi-\theta, -\omega)$. For the square torus, this transformation can be visualized as a rotation by $\pi$ around the origin. For the sphere, it can be visualized as a rotation by $\pi$ through the $x$-axis. We denote this symmetry by $\mathbb{Z}_{2}'$ and the RS1 orbifold symmetry by $\mathbb{Z}_{2}$.

\subsection{7D Clifford Algebra \label{subsec:7DCA}}

The Fermion Kaluza-Klein decomposition in $\textrm{AdS}_{5}\times T_{2}$ has already been examined \cite{McDonald1}. Although our method differs from \cite{McDonald1}, we will ultimately obtain the same result but with added conceptual benefits. Before describing our method and how it differs, though, we will quickly review higher-dimensional Clifford algebras and their representations.

In general, Dirac representations of the $n$-dimensional Clifford algebra have $2^{n/2}$ dimensions if $n$ is even, and $2^{(n-1)/2}$ dimensions if $n$ is odd. One way to represent a general Clifford algebra,
	\[
		\{\Gamma^{M},\Gamma^{N}\}=2\eta^{MN}I
	\]
is by taking a tensor product of lower-dimensional algebras. In the case of the 6D Clifford algebra, the representation is constructed by
	\begin{equation}
		\Gamma^{\mu}=\gamma^{\mu}\otimes \left( -i\sigma^{3} \right), \medspace
		\Gamma^{5/6}=\mathbb{I}\otimes \sigma^{1/2}
		\label{tensorrep}
	\end{equation}
As chiral projections will be very useful for the remainder of the paper, we use the Weyl representation of the 4D Dirac matrices, which is:
	\begin{align}
			\gamma^{0} &=
				\begin{pmatrix}
					0	& 1\\
					1	& 0\\
				\end{pmatrix}, &
			\gamma^{i} &=
				\begin{pmatrix}
					0			& \sigma^{i} \\
					-\sigma^{i}	& 0 \\
				\end{pmatrix}, &
			\gamma^{5} &= \begin{pmatrix}
					-1	& 0\\
					0	& 1\\
				\end{pmatrix}
		\label{eq:4x4matrices}
	\end{align}
where the index $i$ runs from 1 to 3. The first component of each tensor product in eq.~\ref{tensorrep} represents the ordinary 4D Clifford algebra, while the second component describes the Clifford algebra on the internal space. In 2D, the first two Pauli matrices $\sigma^{1/2}$ form a representation of the Clifford algebra, while the third Pauli matrix $\sigma^{3}=-i \sigma^{1}\sigma^{2}$ acts as a chirality projector, similar to $\gamma^{5}$ in 4D.

If we raise the dimension of the internal space to 3, the third Pauli matrix no longer acts as a chirality projector. In general, there are no Weyl fermions in odd dimensions. However, $\sigma^{3}$ instead becomes the final generator of the Clifford algebra. An analogous thing happens in the transition from 6D to 7D regardless of the matrix representation we use. What was the 6D chirality projector, $\Gamma^{7}=i\Gamma^{0}\Gamma^{1}\Gamma^{2}\Gamma^{3}\Gamma^{5}\Gamma^{6}$, becomes the final gamma matrix in 7D. While we can no longer use the operator $\tfrac{1}{2}(1\pm i\Gamma^{7})$ to project onto Weyl fermions in 7D (again, because they do not exist), we can still project out four of the eight components, with the ultimate aim of interpreting the ones that remain as a 4D Dirac spinor.

However, projection via $\Gamma^{7}$ is not the only way to generate a 4-component spinor. We can also project out using the 4D chirality operator $\Gamma^{0}\Gamma^{1}\Gamma^{2}\Gamma^{3}$ or the operator $\Gamma^{5}\Gamma^{6}$. Calculating in the tensor product representation of the 7D Clifford algebra, it is easy to see that $\Gamma^{5}\Gamma^{6}=\mathbb{I}\otimes\sigma^{3}$ is the internal space chirality projector --- or at least, it would be if we were working in 6D. Since the internal space is three-dimensional, this operator doesn't actually project onto a chirality of any subspace of $\textrm{AdS}_{5}\times T_{2}$.

Each of these decompositions can legitimately be reinterpreted as a 4D Dirac spinor. First, however, note that the above representation, while a very useful conceptual tool, is not actually the one we will be using in this paper. So that we can better compare our results with those of McDonald \cite{McDonald1}, we instead use the representation described in that paper, which is:
\begin{align}
		\Gamma^0 &= \begin{pmatrix}
				0	& 1 \\
				1	& 0
			\end{pmatrix} &
		\Gamma^i &= \begin{pmatrix}
				0					& \gamma^0 \gamma^i \\
				- \gamma^0 \gamma^i	& 0
			\end{pmatrix} \nonumber \\
		\Gamma^5 &= i \begin{pmatrix}
				0					& \gamma^0 \gamma^5 \\
				- \gamma^0 \gamma^5	& 0
			\end{pmatrix} &
		\Gamma^6 &= \begin{pmatrix}
				0			& \gamma^0 \\
				- \gamma^0	& 0
			\end{pmatrix}\nonumber \\
		\Gamma^7 &= i \begin{pmatrix}
				-1	& 0 \\
				0	& 1
			\end{pmatrix} \label{ourrep}
	\end{align}
Again, the index $i$ runs from 0 to 3. Note that the change of representation does not modify any of the physics. In particular, it does not change the physical effects of any of the projectors described above, though the 8-component spinor does need to be reordered. The ordering corresponding to the representation in eq.~\ref{ourrep} is
	\[	\Psi=(\Psi_{-R (U)},\Psi_{-L (D)},\Psi_{+L (U)},\Psi_{+R (D)})^{T}	\]
where the decomposition via the 6D chirality operator $\Gamma^{7}$ is notated $\Psi=\Psi_{+}+\Psi_{-}$, the decomposition by the 4D chirality operator $\Gamma^{0}\Gamma^{1}\Gamma^{2}\Gamma^{3}$ is notated $\Psi=\Psi_{R}+\Psi_{L}$, and the decomposition by the internal space chirality operator $\Gamma^{5}\Gamma^{6}$ is notated $\Psi=\Psi_{U}+\Psi_{D}$. The projectors can be explicitly written as 
	\begin{align}
		P_{+/-} 	&= \frac{1}{2}(1\mp i \Gamma^{7})	\nonumber \\
		P_{R/L}	&= \frac{1}{2}(1\pm i\Gamma^{0}\Gamma^{1}\Gamma^{2}\Gamma^{3})
		\label{projectors} \\
		P_{U/D}	&= \frac{1}{2}(1\pm i \Gamma^{5}\Gamma^{6}) \nonumber
	\end{align}
Our Kaluza-Klein decomposition differs from that of McDonald in that we choose $\Psi_{U}$ and $\Psi_{D}$ to represent the four-dimensional Dirac fermions, rather than $\Psi_{+}$ and $\Psi_{-}$. While (as we show) this doesn't change any of the physics, it does provide a useful conceptual advantage. Finally, note first that neither of the two decompositions actually correspond to a 7D-chiral decomposition, as no such decomposition exists. 

During our Kaluza-Klein decomposition, we will identify the 8-component fermions $\psi_{U}$ and $\psi_{D}$ (which have four nonzero components and four zero components) with 4D Dirac spinors $\psi_{U}^{4}$ and $\psi_{D}^{4}$. At this point, we find that the adjoint spinors $\bar{\psi}_{U/D}$=$(\psi_{U/D})^{\dagger} \Gamma_{0}$ and $\bar{\psi}^{4}_{U/D}=(\psi^{4}_{U/D})^{\dagger}\gamma_{0}$ are also identified, since the action of $\Gamma^{0}$ on the 8-component spinors is identical to the action of $\gamma^{0}$ on the 4-component spinors. Specifically, both $\Gamma^{0}$ and $\gamma^{0}$ flip $\psi_{-R}$ with $\psi_{+L}$, and $\psi_{-L}$ with $\psi_{+R}$.

This is not the case if we decompose the spinors by 6D chirality $+/-$. The action of $\Gamma^{0}$ is the same, but $\gamma^{0}$, acting on the four-component spinors, will instead flip $\psi_{-R}$ with $\psi_{+R}$ and $\psi_{-L}$ with $\psi_{+L}$. Working everything out in terms of projectors, we see that this leads to a certain awkwardness in the notation, where $\bar{\psi}_{-R}$ is associated with $\bar{\psi}_{+R}^{4}$, and so on. Decomposing the spinors by the internal space chirality avoids this type of confusion and simplifies calculations.

We still need to show that it is legitimate to associate $\psi_{U/D}$ with 4-component spinors. Usually, the criterion which allows us to identify a subrepresentation is given by Schur's Lemma: the projector onto the subrepresentation must commute with the Lorentz generators $S^{ab}=[\Gamma^{a},\Gamma^{b}]$. Since 7D fermions are not reducible, there will be no projector that satisfies $[P,S^{ab}]=0$ for all $a$ and $b$. However, it can be shown that all three of the projectors described in  eq.~\ref{projectors} commute with $S^{\mu \nu}$, where $\mu$ and $\nu$ run from 0 to 3. This is all that should be required in our decomposition, since we want to reinterpret these as 4D Dirac fermions.

\subsection{Torus \label{subsec:Torus}}

We are now in a position to begin the Kaluza-Klein decomposition of the 7D fermion on $\textrm{AdS}_{5}\times T_{2}$. We begin by writing down an explicit four-component representation of the 4D Dirac fermions, namely $\psi_{U}^{4}=(\psi_{-R},\psi_{+L})^{T}$ and $\psi_{D}^{4}=(\psi_{-L},\psi_{+R})^{T}$ The 4D Dirac matrices $\gamma^{\mu}$ should be understood to act on these spinors, at least when they are not being used to define higher-dimensional Clifford algebras.

We also define the matrices  $\gamma^{\mu}_{8}=\mathbb{I}_{2}\otimes \gamma^{\mu}$, $\sigma^{i}_{8}=\mathbb{I}_{4}\otimes \sigma^{i}$, and $\sigma^{i}_{4}=\mathbb{I}_{2}\otimes \sigma^{i}$, where $\mathbb{I}_{n}$ is the $n\times n$ identity matrix.

Using the explicit representation of the 7D Clifford algebra, we can obtain the following useful relations:
	\begin{align*}
		\gamma^{0}\psi_{+L}^{4} &= \psi_{-R}^{4},
			&\gamma^{0}\psi_{-R}^{4} &=\psi_{+L}^{4} \\
		\gamma^{0}\psi_{-L}^{4} &= \psi_{+R}^{4},
			&\gamma^{0}\psi_{+R}^{4} &= \psi_{-L}^{4} \\
		\gamma^{i}\psi_{+L}^{4} &= \sigma^{i}_{4}\psi_{-R}^{4},
			&\gamma^{i}\psi_{-R}^{4} &= -\sigma^{i}_{4}\psi_{+L}^{4} \\
		\gamma^{i}\psi_{-L}^{4} &= -\sigma^{i}_{4}\psi_{+R}^{4},
			&\gamma^{i}\psi_{+R}^{4} &= \sigma^{i}_{4}\psi_{-L}^{4}
	\end{align*}
We indicate this reduction of an 8-component fermion to a 4-component fermion with an arrow, $\rightarrow$. We can now calculate terms of the following form, which will appear in the Kaluza-Klein decomposition.
	\begin{align}
		\Gamma^{0}\psi_{+R} =\psi_{-L}&\rightarrow \psi_{-L}^{4}=\gamma^{0}\psi_{+R}^{4}\notag\\
		\Gamma^{0}\psi_{+L} =\psi_{-R}&\rightarrow \psi_{-R}^{4}=\gamma^{0}\psi_{+L}^{4}\notag\\
		\Gamma^{0}\psi_{-R} =\psi_{+L}&\rightarrow \psi_{+L}^{4}=\gamma^{0}\psi_{-R}^{4}\notag\\
		\Gamma^{0}\psi_{-L} =\psi_{+R}&\rightarrow \psi_{+R}^{4}=\gamma^{0}\psi_{-L}^{4}
		\label{Gamma 0 Terms}
	\end{align}
We also have:
	\begin{align}
		\Gamma^{i}\psi_{+R}=\gamma^{0}_{8}\gamma^{i}_{8}\psi_{-L}=\sigma^{i}_{8}\psi_{-L}&\rightarrow \sigma^{i}_{4}\psi_{-L}^{4}
			=\gamma^{i}\psi_{+R}^{4}\notag\\
		\Gamma^{i}\psi_{+L}=\gamma^{0}_{8}\gamma^{i}_{8}\psi_{-R}=-\sigma^{i}_{8}\psi_{-R}&\rightarrow -\sigma^{i}_{4}\psi_{-R}^{4}=-\gamma^{i}
			\psi_{+L}^{4}\notag\\
		\Gamma^{i}\psi_{-R}=-\gamma^{0}_{8}\gamma^{i}_{8}\psi_{+L}=\sigma^{i}_{8}\psi_{+L}&\rightarrow \sigma^{i}_{4}\psi_{+L}^{4}=-\gamma^{i}
			\psi_{-R}^{4}\notag\\
		\Gamma^{i}\psi_{-L}=-\gamma^{0}_{8}\gamma^{i}_{8}\psi_{+R}=-\sigma^{i}_{8}\psi_{+R}&\rightarrow -\sigma^{i}_{4}\psi_{+R}^{4}=\gamma^{i}
			\psi_{-L}^{4}
		\label{Gamma Mu Terms}
	\end{align}
We can learn even more about terms of the form $\Gamma^{5/6}\psi$.
	\begin{align}
		\Gamma^{5}\psi_{+R}=i\gamma^{0}_{8}\gamma^{5}_{8}\psi_{-L}=i\psi_{-R}&\rightarrow i\psi^{4}_{-R}\notag\\
		\Gamma^{5}\psi_{+L}=i\gamma^{0}_{8}\gamma^{5}_{8}\psi_{-R}=-i\psi_{-L}&\rightarrow -i\psi^{4}_{-L}\notag\\
		\Gamma^{5}\psi_{-R}=-i\gamma^{0}_{8}\gamma^{5}_{8}\psi_{+L}=i\psi_{+R}&\rightarrow i\psi^{4}_{+R}\notag\\
		\Gamma^{5}\psi_{-L}=-i\gamma^{0}_{8}\gamma^{5}_{8}\psi_{+R}=-i\psi_{+L}&\rightarrow -i\psi^{4}_{+L}
		\label{Gamma 5 Terms}
	\end{align}
and
	\begin{align}
		\Gamma^{6}\psi_{+R}=\gamma^{0}_{8}\psi_{-L}=\psi_{-R}&\rightarrow\psi^{4}_{-R}\notag\\
		\Gamma^{6}\psi_{+L}=\gamma^{0}_{8}\psi_{-R}=\psi_{-L}&\rightarrow\psi^{4}_{-L}\notag\\
		\Gamma^{6}\psi_{-R}=-\gamma^{0}_{8}\psi_{+L}=-\psi_{+R}&\rightarrow -\psi^{4}_{+R}\notag\\
		\Gamma^{6}\psi_{-L}=-\gamma^{0}_{8}\psi_{+R}=-\psi_{+L}&\rightarrow-\psi^{4}_{+L}
		\label{Gamma 6 Terms}
	\end{align}
Finally, note that $\Gamma^{7}\psi_{-R/L}=-i\psi_{-R/L}$ and $\Gamma^{7}\psi_{+R/L}=i\psi_{+R/L}$.

We are now in a position to begin the Kaluza-Klein decomposition for $\textrm{AdS}_{5}/\mathbb{Z}_{2}\times T_{2}/\mathbb{Z}_{2}'$. The seven-dimensional fermion action is:
	\begin{multline}
		S_{7\textrm{D}}=\int \textrm{d}^{4}x \int \textrm{d} \phi \int \textrm{d} \theta_{1} \textrm{d} \theta_{2}\sqrt{G}\bigg[\frac{i}{2}\bar{\Psi} 
		E^{A}_{a}\Gamma^{a}\partial_{A}\Psi\\
		-\frac{i}{2}(\partial_{A}\bar{\Psi})E^{A}_{a}\Gamma^{a}\Psi-M_{7}\epsilon(\phi)\bar{\Psi}\Psi\bigg]
		\label{Torus Action}
	\end{multline}
where $\sqrt{G}=e^{-4\sigma}r_{c}R^{2}$ and
	\[	E_{a}^{A}=\textrm{diag}\left(e^{\sigma},e^{\sigma},e^{\sigma},e^{\sigma},\frac{1}{r_{c}},\frac{1}{R},\frac{1}{R}\right)	\]
Then, the action becomes
	\begin{multline}
		S_{7\textrm{D}}=\int \textrm{d}^{7}x\bigg[ e^{-3\sigma}r_{c}R^{2}\frac{i}{2}\bar{\Psi}\Gamma^{\mu}\partial_{\mu}\Psi\\
		-e^{-3\sigma}r_{c}R^{2}\frac{i}{2}(\partial_{\mu}\bar{\Psi})\Gamma^{\mu}\Psi +e^{-4\sigma}R^{2}\frac{i}{2}\bar{\Psi}\Gamma^{7}\partial_{7}\Psi\\
		-e^{-4\sigma}R^{2}\frac{i}{2}(\partial_{7}\bar{\Psi})\Gamma^{7}\Psi+e^{-4\sigma}r_{c}R\frac{i}{2}\bar{\Psi}\Gamma^{b}\partial_{b}\Psi-\\
		e^{-4\sigma}r_{c}R\frac{i}{2}(\partial_{b}\bar{\Psi})\Gamma^{b}\Psi-e^{-4\sigma}r_{c}R^{2}M_{7}\epsilon(\phi)\bar{\Psi}\Psi\bigg]
	\end{multline}
where the index $b$ runs over 5 and 6. Integrating by parts to remove the derivatives from the adjoint spinors, we find
	\begin{multline}
		S_{7\textrm{D}}=\int \textrm{d}^{7}x\bigg[e^{-3\sigma}r_{c}R^{2}i\bar{\Psi}\Gamma^{\mu}\partial_{\mu}\Psi\\ +e^{-4\sigma}R^{2}i\bar{\Psi}
		\Gamma^{7}\partial_{7}\Psi - 2kr_{c}\epsilon(\phi)e^{-4\sigma}R^{2}i \bar{\Psi}\Gamma^{7}\Psi\\
		+e^{-4\sigma}r_{c}R i \bar{\Psi}\Gamma^{b}\partial_{b}\Psi-e^{-4\sigma}r_{c}R^{2}M_{7}\epsilon(\phi)\bar{\Psi}\Psi\bigg]
		\label{Simplified Torus Action}
	\end{multline}
At this point, we need to decompose the 7D fermion into Kaluza-Klein modes. The decomposition is
	\begin{multline}
		\Psi=\sum_{m,p_{1},p_{2}}\psi^{m,p}_{-R}f^{m,p}_{-R}g^{p}_{-R}+\psi^{m,p}_{-L}f^{m,p}_{-L}g^{p}_{-L}\\
		+\psi^{m,p}_{+L}f^{m,p}_{+L}g^{p}_{+L}+\psi^{m,p}_{+R}f^{m,p}_{+R}g^{p}_{+R}
		\label{7D KK Decomposition}
	\end{multline}
where $p$ stands for $p_{1}$ and $p_{2}$, $\psi$ is a spinor function depending only on $x^{\mu}$, $f$ is a scalar function depending only on $\phi$ (the Randall-Sundrum warped dimension), and $g$ is a scalar function depending only on $\theta_{1}$ and $\theta_{2}$ (the two dimensions compactified on the torus).

We first simplify the $\partial_{\mu}$ terms. We can reduce these terms to four-component spinors using the gamma-matrix identities we derived earlier. Then, we cancel three-fourths of the resulting terms by using the fact that 4D gamma matrices flip both chiralities. Carrying out this calculation, we find that $\bar{\Psi}\Gamma^{\mu}\partial_{\mu}\Psi$ reduces to 4-component spinors as
	\begin{align}
		&\bar{\Psi}\Gamma^{\mu}\partial_{\mu}\Psi\rightarrow \left(\sum_{m,p}\bar{\psi}_{+R}\bar{f}_{+R}\bar{g}_{+R}\right)\partial_{0}\left(\sum_{n,q}
			\psi_{-L}f_{+R}g_{+R}\right) \nonumber \\
		&\quad +\left(\sum_{m,p}\bar{\psi}_{+L}\bar{f}_{+L}\bar{g}_{+L}\right)\partial_{0}\left(\sum_{n,q}\psi_{-R}f_{+L}g_{+L}\right) \nonumber \\
		&\quad +\left(\sum_{m,p}\bar{\psi}_{-R}\bar{f}_{-R}\bar{g}_{-R}\right)\partial_{0}\left(\sum_{n,q}\psi_{+L}f_{-R}g_{-R}\right) \nonumber \\
		&\quad +\left(\sum_{m,p}\bar{\psi}_{-L}\bar{f}_{-L}\bar{g}_{-L}\right)\partial_{0}\left(\sum_{n,q}\psi_{+R}f_{-L}g_{-L}\right) \nonumber \\
		&\quad +\left(\sum_{m,p}\bar{\psi}_{+R}\bar{f}_{+R}\bar{g}_{+R}\right)\sigma^{i}_{4}\partial_{i}\left(\sum_{n,q}\psi_{-L}f_{+R}g_{+R}\right) 
			\nonumber \\
		&\quad +\left(\sum_{m,p}\bar{\psi}_{+L}\bar{f}_{+L}\bar{g}_{+L}\right)\sigma^{i}_{4}\partial_{i}
			\left(\sum_{n,q}\psi_{-R}f_{+L}g_{+L}\right) \nonumber \\
		&\quad -\left(\sum_{m,p}\bar{\psi}_{-R}\bar{f}_{-R}\bar{g}_{-R}\right)\sigma^{i}_{4}\partial_{i}\left(\sum_{n,q}\psi_{+L}f_{-R}g_{-R}\right) 
			\nonumber \\
		&\quad -\left(\sum_{m,p}\bar{\psi}_{-L}\bar{f}_{-L}\bar{g}_{-L}\right)\sigma^{i}_{4}\partial_{i}\left(\sum_{n,q}\psi_{+R}f_{-L}g_{-L}\right)
	\end{align}
(Here, and for the remainder of the paper, we suppress the superscript-four notation that indicates a reduction to 4-component fermions.)

However, if we expand out the four-component spinors $\bar{\Psi}_{U}\gamma^{\mu}\partial_{\mu}\Psi_{U}+\bar{\Psi}_{D}\gamma^{\mu}\partial_{\mu}\Psi_{D}$ in the same manner, the result is exactly the same! Therefore, we can use this expression to replace $\bar{\Psi}\Gamma^{\mu}\partial_{\mu}\Psi$ in the action. 

Making significant use of the gamma matrix identities (eqs.~\ref{Gamma 0 Terms}--\ref{Gamma 6 Terms}) and eliminating terms where possible, the 7D action, eq.~\ref{Simplified Torus Action}, can now be reduced from 8-component to 4-component spinors. The result is:
	\begin{multline}
		S_{7\textrm{D}}=\int\textrm{d}^{7}x\Bigg(e^{-3\sigma}r_{c}R^{2}
		i\left[\bar{\Psi}_{U}\gamma^{\mu}\partial_{\mu}\Psi_{U}+\bar{\Psi}_{D}\gamma^{\mu}
		\partial_{\mu}\Psi_{D}\right]\\
		+e^{-4\sigma}R^{2}\times\\
		\bigg[\left(\sum_{m,p}\bar{\psi}_{+R}\bar{f}_{+R}\bar{g}_{+R}\right)\partial_{7}\left(\sum_{n,q}\psi_{-L}f_{-L}g_{-L}\right)\\
		+\left(\sum_{m,p}\bar{\psi}_{+L}\bar{f}_{+L}\bar{g}_{+L}\right) \partial_{7}\left(\sum_{n,q}\psi_{-R}f_{-R}g_{-R}\right)\\
		-\left(\sum_{m,p}\bar{\psi}_{-R}\bar{f}_{-R}\bar{g}_{-R}\right)\partial_{7}\left(\sum_{n,q}\psi_{+L}f_{+L}g_{+L}\right)\\
		-\left(\sum_{m,p}\bar{\psi}_{-L}\bar{f}_{-L}\bar{g}_{-L}\right) \partial_{7}\left(\sum_{n,q}\psi_{+R}f_{+R}g_{+R}\right)\bigg]\\
		-2kr_{c}\epsilon(\phi)e^{-4\sigma}R^{2}\times\\
		\bigg[\left(\sum_{m,p}\bar{\psi}_{+R}\bar{f}_{+R}\bar{g}_{+R}\right)\left(\sum_{n,q}\psi_{-L}f_{-L}g_{-L}\right)\\
		+\left(\sum_{m,p}\bar{\psi}_{+L}\bar{f}_{+L}\bar{g}_{+L}\right)\left(\sum_{n,q}\psi_{-R}f_{-R}g_{-R}\right)\\
		-\left(\sum_{m,p}\bar{\psi}_{-R}\bar{f}_{-R}\bar{g}_{-R}\right)\left(\sum_{n,q}\psi_{+L}f_{+L}g_{+L}\right)\\
		-\left(\sum_{m,p}\bar{\psi}_{-L}\bar{f}_{-L}\bar{g}_{-L}\right)\left(\sum_{n,q}\psi_{+R}f_{+R}g_{+R}\right)\bigg]\\
		+e^{-4\sigma}r_{c}R\times\\
		\bigg[\left(\sum_{m,p}\bar{\psi}_{+R}\bar{f}_{+R}\bar{g}_{+R}\right)(\partial_{5}+i\partial_{6})\left(\sum_{n,q}\psi_{-L}f_{+L}g_{+L}\right)\\
		+\left(\sum_{m,p}\bar{\psi}_{+L}\bar{f}_{+L}\bar{g}_{+L}\right)(-\partial_{5}+i\partial_{6})\left(\sum_{n,q}\psi_{-R}f_{+R}g_{+R}\right)\\
		+\left(\sum_{m,p}\bar{\psi}_{-R}\bar{f}_{-R}\bar{g}_{-R}\right)(\partial_{5}-i\partial_{6})\left(\sum_{n,q}\psi_{+L}f_{-L}g_{-L}\right)\\
		+\left(\sum_{m,p}\bar{\psi}_{-L}\bar{f}_{-L}\bar{g}_{-L}\right)(-\partial_{5}-i\partial_{6})\left(\sum_{n,q}\psi_{+R}f_{-R}g_{-R}\right)\bigg]\\
		-e^{-4\sigma}r_{c}R^{2}M_{7}\epsilon(\phi)\times\\
		\bigg[\left(\sum_{m,p}\bar{\psi}_{+R}\bar{f}_{+R}\bar{g}_{+R}\right)\left(\sum_{n,q}\psi_{-L}f_{-L}g_{-L}\right)\\
		+\left(\sum_{m,p}\bar{\psi}_{+L}\bar{f}_{+L}\bar{g}_{+L}\right)\left(\sum_{n,q}\psi_{-R}f_{-R}g_{-R}\right)\\+\left(\sum_{m,p}\bar{\psi}_{-R}
		\bar{f}_{-R}\bar{g}_{-R}\right)\left(\sum_{n,q}\psi_{+L}f_{+L}g_{+L}\right)\\
		+\left(\sum_{m,p}\bar{\psi}_{-L}\bar{f}_{-L}\bar{g}_{-L}\right)\left(\sum_{n,q}\psi_{+R}f_{+R}g_{+R}\right)\bigg]\Bigg)
		\label{Big Torus Action}
	\end{multline}
We can now use the fact that each term in this action must be even under $\mathbb{Z}_{2}$ and $\mathbb{Z}_{2}'$ parity transformations to determine the parities of the different spinor components. We begin with $\mathbb{Z}_{2}$ parity. The $\partial_{\mu}$ terms tell us nothing, since the only surviving terms will look like $\bar{f}_{+L}f_{+L}$, and these terms are always even. From the $\partial_{4}$ terms, we see that $f_{+R}$ and $f_{-L}$ must have opposite parity, and $f_{-R}$ and $f_{+L}$ must have opposite parity. This follows from the fact that $\partial_{4}$ is $\mathbb{Z}_{2}$-odd. Meanwhile, the $\partial_{5}$ and $\partial_{6}$ terms tell us that $f_{+R}$ and $f_{+L}$ have the same parity, and that $f_{-R}$ and $f_{-L}$ have the same parity. The remaining terms agree with these conclusions. 

Now, we examine everything under $\mathbb{Z}_{2}'$ parity. From the $\partial_{4}$ terms, we learn that $g_{+R}$ and $g_{-L}$ have the same parity, and that $g_{+L}$ and $g_{-R}$ have the same parity. From the $\partial_{5}$ and $\partial_{6}$ terms, we learn that $g_{+R}$ and $g_{+L}$ have opposite parity, and that $g_{-R}$ and $g_{-L}$ have opposite parity. As before, the remaining terms agree.

A parity choice that agrees with all these specifications is:
	\begin{equation}
		\Psi=\left(\begin{array}{c}
		\Psi_{-R}: (-,+)\\
		\Psi_{-L}: (-,-)\\
		\Psi_{+L}: (+,+)\\
		\Psi_{+R}: (+,-)
		\end{array}\right)
		\label{Parities}
	\end{equation}
We can now begin the Kaluza-Klein decomposition proper. We compare the above action to the RS1 fermion action (eq.~\ref{eq:RS1action}) which, through a similar but much simpler procedure, can be written as:
	\begin{multline}
		S_{5\textrm{D}}=\int \textrm{d}^{5}x\Bigg(e^{-3\sigma}r_{c}i\bigg[\sum_{m}\sum_{n}\bar{f}^{m}_{A}\bar{\psi}^{m}_{A}\gamma^{\mu}
		\partial_{\mu}f^{n}_{A}\psi^{n}_{A}\\
		+\sum_{m}\sum_{n}\bar{f}^{m}_{B}\bar{\psi}^{m}_{B}\gamma^{\mu}\partial_{\mu}f^{n}_{B}\psi^{n}_{B}\bigg]
		-e^{-4\sigma}\bigg[\sum_{m}\sum_{n}\bar{f}^{m}_{A}\bar{\psi}^{m}_{A}\partial_{7}f^{n}_{B}\psi^{n}_{B}\\
		-\sum_{m}\sum_{n}\bar{f}^{m}_{B}\bar{\psi}^{m}_{B}\partial_{7}f^{n}_{A}\psi^{n}_{A}\bigg]
		+2kr_{c}\epsilon(\phi)e^{-4\sigma}\\
		\bigg[\sum_{m}\sum_{n}\bar{f}^{m}_{A}\bar{\psi}^{m}_{A}f^{n}_{B}\psi^{n}_{B}
		-\sum_{m}\sum_{n}\bar{f}^{m}_{B}\bar{\psi}^{m}_{B}f^{n}_{A}\psi^{n}_{A}\bigg]-Mr_{c}\epsilon(\phi)e^{-4\sigma}\\
		\bigg[\sum_{m}\sum_{n}\bar{f}^{m}_{A}\bar{\psi}^{m}_{A}f^{n}_{B}\psi^{n}_{B}
		+\sum_{m}\sum_{n}\bar{f}^{m}_{B}\bar{\psi}^{m}_{B}f^{n}_{A}\psi^{n}_{A}\bigg]\Bigg)
		\label{Expanded RS1 Action}
	\end{multline}
where $A$ and $B$ are two chiralities. Note that we will need two copies of this formula (and a double sum over KK modes) to fully represent our reduced 7D fermion. Comparing 4D kinetic terms, we obtain the normalization condition:
	\[
		\int_{-\pi}^{\pi}\int_{-\pi}^{\pi}\bar{g}_{+L}^{p_{1},p_{2}},g_{+L}^{q_{1},q_{2}}\textrm{d}\theta_{1}\textrm{d}\theta_{2}=\delta_{p_{1}q_{1}}
		\delta_{p_{2}q_{2}}
	\]
and identical formulas for the other chiralities. Comparing the next two terms in each formula yields $g_{-L}=g_{+R}$ and $g_{-R}=g_{+L}$. Finally, requiring the $\partial_{5}$ and $\partial_{6}$ terms in the 7D action to integrate to yield a 5D mass term produces the differential equations
	\begin{align}
		(\partial_{5}-i\partial_{6})g_{-L}^{p_{1},p_{2}}=&M_{p_{1},p_{2}}g_{-R}^{p_{1},p_{2}}\label{Torus DEQ 1}\\
		(\partial_{5}+i\partial_{6})g_{-R}^{p_{1},p_{2}}=&-M_{p_{1},p_{2}}g_{-L}^{p_{1},p_{2}}
		\label{Torus DEQ 2}
	\end{align}
We now define
	\[
		\xi^{p_{1},p_{2}}_{\textrm{A}}=\sum_{m}\psi^{m,p_{1},p_{2}}_{A} f^{m,p_{1},p_{2}}_{A}
	\]
where $A$ is an arbitrary chirality. Finally the 7D action, eq.~\ref{Big Torus Action}, reduces to five dimensions:
	\begin{multline}
		S_{5D}=\int\textrm{d}^{5}x\Bigg(e^{-3\sigma}r_{c}i\sum_{p_{1},p_{2}}\bigg[\bar{\xi}_{U}\gamma^{\mu}\partial_{\mu}\xi_{U}+\bar{\xi}_{D}
		\gamma^{\mu}\partial_{\mu}\xi_{D}\bigg]\\
		-e^{-4\sigma}\sum_{p_{1},p_{2}}\bigg[\bar{\xi}_{-R}\partial_{7}\xi_{+L}-\bar{\xi}_{+L}\partial_{7}\xi_{-R}+\bar{\xi}_{-L}\partial_{7}\xi_{+R}\\
		-\bar{\xi}_{+R}\partial_{7}\xi_{-L}\bigg]+2kr_{c}\epsilon(\phi)e^{-4\sigma}\sum_{p_{1},p_{2}}\bigg[\bar{\xi}_{-R}\xi_{+L}-\bar{\xi}_{+L}\xi_{-R}\\
		+\bar{\xi}_{-L}\xi_{+R}-\bar{\xi}_{+R}\xi_{-L}\bigg]+e^{-4\sigma}\frac{r_{c}}{R}\sum_{p_{1},p_{2}}\Bigg[M_{p_{1},p_{2}}\sum_{m}\sum_{n}\\
		\big(\bar{\psi}_{-R}\bar{f}_{-R}f_{-L}\psi_{+L}-\bar{\psi}_{+L}\bar{f}_{+L}f_{+R}\psi_{-R}
		+\bar{\psi}_{-L}\bar{f}_{-L}f_{-R}\psi_{+R}\\
		-\bar{\psi}_{+R}\bar{f}_{+R}f_{+L}\psi_{-L}\big)\Bigg]
		-e^{-4\sigma}r_{c}M_{7}\epsilon(\phi)\sum_{p_{1},p_{2}}\bigg[\bar{\xi}_{-R}\xi_{+L}\\
		+\bar{\xi}_{+L}\xi_{-R}+\bar{\xi}_{-L}\xi_{+R}+\bar{\xi}_{+R}\xi_{-L}\bigg]\Bigg)
		\label{5D Reduced Action}
	\end{multline}
Before reducing our 5D action to four dimensions, let's solve the coupled differential equations in eqs.~\ref{Torus DEQ 1} and \ref{Torus DEQ 2}. The eigenvalues are $M_{p_{1},p_{2}}=\sqrt{p_{1}^{2}+p_{2}^{2}}$. The even-parity solutions are
	\[	g_{-R}=g_{+L}\propto \cos(p_{1}\theta_{1}+p_{2}\theta_{2})	\]
and the odd-parity solutions are
	\[	g_{-L}=g_{+R}\propto \frac{p_{1}+ip_{2}}{\sqrt{p_{1}^{2}+p_{2}^{2}}}\sin(p_{1}\theta_{1}+p_{2}\theta_{2}) 	\]
The factor in the odd-parity solutions ensures that the two proportionality constants for these solutions are equal.

Continuing with the KK reduction, we compare eq.~\ref{5D Reduced Action} to two copies of the following triple sum over 4D fermion actions:
	\begin{multline}
		S_{4\textrm{D}}=\int \textrm{d}^{4}x\sum_{m}\sum_{p_{1},p_{2}}\big(i\bar{\psi}_{A}\gamma^{\mu}\partial_{\mu}\psi_{A}+i\bar{\psi}_{B}
		\gamma^{\mu}\partial_{\mu}\psi_{B}\\
		-m\bar{\psi}_{A}\psi_{B}-m\bar{\psi}_{B}\psi_{A}\big)
	\end{multline}
Comparing just the kinetic terms, we find the following normalization condition:
	\[	\int_{-\pi}^{\pi} e^{-3\sigma}r_{c}\bar{f}^{m}_{+L}f^{n}_{+L}\textrm{d}\phi=\delta_{mn}	\]
and identical formulas for the other chiralities. Requiring that the remaining terms in the 5D action take the form of 4D mass terms after integration yields a set of four coupled differential equations:
	\begin{multline}
		\frac{-e^{-\sigma}}{r_{c}}\partial_{7}f_{+L}+2k\epsilon(\phi)e^{-\sigma}f_{+L}+\frac{e^{-\sigma}}{R}M_{p_{1},p_{2}}f_{-L}\\
		-e^{-\sigma}M_{7}\epsilon(\phi)f_{+L}=-mf_{-R}\\\
		\frac{e^{-\sigma}}{r_{c}}\partial_{7}f_{-R}-2k\epsilon(\phi)e^{-\sigma}f_{-R}-\frac{e^{-\sigma}}{R}M_{p_{1},p_{2}}f_{+R}\\
		-e^{-\sigma}M_{7}\epsilon(\phi)f_{-R}=-mf_{+L}\\\
		\frac{-e^{-\sigma}}{r_{c}}\partial_{7}f_{+R}+2k\epsilon(\phi)e^{-\sigma}f_{+R}+\frac{e^{-\sigma}}{R}M_{p_{1},p_{2}}f_{-R}\\
		-e^{-\sigma}M_{7}\epsilon(\phi)f_{+R}=-mf_{-L}\\\
		\frac{e^{-\sigma}}{r_{c}}\partial_{7}f_{-L}-2k\epsilon(\phi)e^{-\sigma}f_{-L}-\frac{e^{-\sigma}}{R}M_{p_{1},p_{2}}f_{+L}\\
		-e^{-\sigma}M_{7}\epsilon(\phi)f_{-L}=-mf_{+R}
	\end{multline}
We can further simplify these equations with the substitution $\hat{f}=e^{-2\sigma}f$, giving the following:
	\begin{multline}
		\frac{-e^{-\sigma}}{r_{c}}\partial_{7}\hat{f}_{+L}+\frac{e^{-\sigma}}{R}M_{p_{1},p_{2}}\hat{f}_{-L}-e^{-\sigma}M_{7}\epsilon(\phi)\hat{f}_{+L}
			=-m\hat{f}_{-R}\\
		\frac{e^{-\sigma}}{r_{c}}\partial_{7}\hat{f}_{-R}-\frac{e^{-\sigma}}{R}M_{p_{1},p_{2}}\hat{f}_{+R}-e^{-\sigma}M_{7}\epsilon(\phi)\hat{f}_{-R}
			=-m\hat{f}_{+L}\\
		\frac{-e^{-\sigma}}{r_{c}}\partial_{7}\hat{f}_{+R}+\frac{e^{-\sigma}}{R}M_{p_{1},p_{2}}\hat{f}_{-R}-e^{-\sigma}M_{7}\epsilon(\phi)\hat{f}_{+R}
			=-m\hat{f}_{-L}\\
		\frac{e^{-\sigma}}{r_{c}}\partial_{7}\hat{f}_{-L}-\frac{e^{-\sigma}}{R}M_{p_{1},p_{2}}\hat{f}_{+L}-e^{-\sigma}M_{7}\epsilon(\phi)\hat{f}_{-L}
			=-m\hat{f}_{+R}
		\label{7D DEQ 4}
	\end{multline}
As in the RS1 case, these equations are solved on $[0,\pi)$ by applying homogeneous Dirichlet boundary conditions for the odd modes and nonhomogeneous Neumann boundary conditions for the even modes. The former determines the eigenvalues. Similiar to \cite{McDonald1}, we obtain a chiral zero-mode, which, for our parity choice, is $\Psi_{+L}$.

\subsection{Sphere \label{subsec:Sphere}}

We now move on to $\textrm{AdS}_{5}/\mathbb{Z}_{2}\times S_{2}/\mathbb{Z}_{2}'$. There are many similarities between the sphere and the torus, with a few important differences. The 7D fermion action has the same initial form as above (eq.~\ref{Torus Action}). However,
	\[	\sqrt{G}=e^{-4\sigma}r_{c}R^{2}\sin\theta	\]
and
	\[	E_{a}^{A}=\textrm{diag}\left(e^{\sigma},e^{\sigma},e^{\sigma},e^{\sigma},\frac{1}{r_{c}},\frac{1}{R},\frac{1}{R\sin\theta}\right)	\]
Plugging these terms in, the 7D action becomes:
	\begin{multline}
		S_{7\textrm{D}}=\int\textrm{d}^{7}x\bigg[  \frac{i}{2} e^{-3\sigma}r_{c}R^{2}\sin \theta \left(\bar{\Psi}\Gamma^{\mu}\partial_{\mu}\Psi
		-\left(\partial_{\mu}\bar{\Psi}\right)\Gamma^{\mu}\Psi\right)\\
		+ \frac{i}{2} e^{-4\sigma}R^{2}\sin \theta
		\left(\bar{\Psi}\Gamma^{7}\partial_{7}\Psi-\left(\partial_{7}\bar{\Psi}\right)\Gamma^{7}\Psi\right)\\
		+ \frac{i}{2} e^{-4\sigma}r_{c}R\sin \theta \left(\bar{\Psi}\Gamma^{5}\partial_{5}\Psi-\left(\partial_{5}\bar{\Psi}\right)\Gamma^{5}\Psi\right)\\
		+ \frac{i}{2} e^{-4\sigma}r_{c}R \left(\bar{\Psi}\Gamma^{6}\partial_{6}\Psi-\left(\partial_{6}\bar{\Psi}\right)\Gamma^{6}\Psi\right)\\
		- e^{-4\sigma}r_{c}R^{2}\sin\theta M_{7}\epsilon(\phi)\bar{\Psi}\Psi\bigg]
	\end{multline}
After integration by parts, the action becomes
	\begin{multline}
		S_{7\textrm{D}}=\int\textrm{d}^{7}x\bigg[i e^{-3\sigma}r_{c}R^{2}\sin \theta \bar{\Psi}\Gamma^{\mu}\partial_{\mu}\Psi\\
		+ i \,e^{-4\sigma}R^{2}\sin \theta \bar{\Psi}\Gamma^{7}\partial_{7}\Psi
		- 2i \, k r_{c}\epsilon (\phi)e^{-4\sigma}R^{2}\sin \theta \bar{\Psi}\Gamma^{7}\Psi\\
		+ i \,e^{-4\sigma}r_{c}R\sin \theta \bar{\Psi}\Gamma^{5}\partial_{5}\Psi
		+ \frac{i}{2}e^{-4\sigma}r_{c}R\cos \theta  \bar{\Psi}\Gamma^{5}\Psi\\
		+ i \, e^{-4\sigma}r_{c}R \bar{\Psi}\Gamma^{6}\partial_{6}\Psi-e^{-4\sigma}r_{c}R^{2}\sin\theta M_{7}\epsilon(\phi)\bar{\Psi}\Psi\bigg]
	\end{multline}
Using the machinery we developed for the torus, we can now apply the Kaluza-Klein decomposition and reduce the action to four-component spinors. The result is the following:
	\begin{multline}
		S_{7\textrm{D}}=\int\textrm{d}^{7}x\Bigg( i \, e^{-3\sigma}r_{c}R^{2}\sin \theta\\
			\left[\bar{\Psi}_{U}\gamma^{\mu}\partial_{\mu}\Psi_{U}+\bar{\Psi}_{D}\gamma^{\mu}
			\partial_{\mu}\Psi_{D}\right]\\
		+e^{-4\sigma}R^{2}\sin \theta \times\\
		\bigg[\left(\sum_{m,p}\bar{\psi}_{+R}\bar{f}_{+R}\bar{g}_{+R}\right)\partial_{7}\left(\sum_{n,q}\psi_{-L}f_{-L}g_{-L}\right)\\
			+\left(\sum_{m,p}\bar{\psi}_{+L}\bar{f}_{+L}\bar{g}_{+L}\right)\partial_{7}\left(\sum_{n,q}\psi_{-R}f_{-R}g_{-R}\right)\\
			-\left(\sum_{m,p}\bar{\psi}_{-R}\bar{f}_{-R}\bar{g}_{-R}\right)\partial_{7}\left(\sum_{n,q}\psi_{+L}f_{+L}g_{+L}\right)\\
			-\left(\sum_{m,p}\bar{\psi}_{-L}\bar{f}_{-L}\bar{g}_{-L}\right)\partial_{7}\left(\sum_{n,q}\psi_{+R}f_{+R}g_{+R}\right)\bigg]\\
		-2kr_{c}\epsilon(\phi)e^{-4\sigma}R^{2}\sin \theta \times\\
		 \bigg[\left(\sum_{m,p}\bar{\psi}_{+R}\bar{f}_{+R}\bar{g}_{+R}\right)
			\left(\sum_{n,q}\psi_{-L}f_{-L}g_{-L}\right)\\
			+\left(\sum_{m,p}\bar{\psi}_{+L}\bar{f}_{+L}\bar{g}_{+L}\right)\left(\sum_{n,q}\psi_{-R}f_{-R}g_{-R}\right)\\
			-\left(\sum_{m,p}\bar{\psi}_{-R}\bar{f}_{-R}\bar{g}_{-R}\right)\left(\sum_{n,q}\psi_{+L}f_{+L}g_{+L}\right)\\
			-\left(\sum_{m,p}\bar{\psi}_{-L}\bar{f}_{-L}\bar{g}_{-L}\right)\left(\sum_{n,q}\psi_{+R}f_{+R}g_{+R}\right)\bigg]\\
		+e^{-4\sigma}r_{c}R\sin \theta \times \\
		\bigg[\left(\sum_{m,p}\bar{\psi}_{+R}\bar{f}_{+R}\bar{g}_{+R}\right)\partial_{5}\left(\sum_{n,q}\psi_{-L}f_{+L}g_{+L}\right)\\
			-\left(\sum_{m,p}\bar{\psi}_{+L}\bar{f}_{+L}\bar{g}_{+L}\right)\partial_{5}\left(\sum_{n,q}\psi_{-R}f_{+R}g_{+R}\right)\\
			+\left(\sum_{m,p}\bar{\psi}_{-R}\bar{f}_{-R}\bar{g}_{-R}\right)\partial_{5}\left(\sum_{n,q}\psi_{+L}f_{-L}g_{-L}\right)\\
			-\left(\sum_{m,p}\bar{\psi}_{-L}\bar{f}_{-L}\bar{g}_{-L}\right)\partial_{5}\left(\sum_{n,q}\psi_{+R}f_{-R}g_{-R}\right)\bigg]\\
		+\frac{1}{2}e^{-4\sigma}r_{c}R\cos \theta \times\\
		\bigg[\left(\sum_{m,p}\bar{\psi}_{+R}\bar{f}_{+R}\bar{g}_{+R}\right)\left(\sum_{n,q}\psi_{-L}f_{+L}g_{+L}\right)\\
			-\left(\sum_{m,p}\bar{\psi}_{+L}\bar{f}_{+L}\bar{g}_{+L}\right)\left(\sum_{n,q}\psi_{-R}f_{+R}g_{+R}\right)\\
			+\left(\sum_{m,p}\bar{\psi}_{-R}\bar{f}_{-R}\bar{g}_{-R}\right)\left(\sum_{n,q}\psi_{+L}f_{-L}g_{-L}\right)\\
			-\left(\sum_{m,p}\bar{\psi}_{-L}\bar{f}_{-L}\bar{g}_{-L}\right)\left(\sum_{n,q}\psi_{+R}f_{-R}g_{-R}\right)\bigg]\\
		+ i \, e^{-4\sigma}r_{c}R \times\\ 
			\bigg[\left(\sum_{m,p}\bar{\psi}_{+R}\bar{f}_{+R}\bar{g}_{+R}\right)\partial_{6}\left(\sum_{n,q} \psi_{-L}f_{+L}g_{+L}\right)\\
			+\left(\sum_{m,p}\bar{\psi}_{+L}\bar{f}_{+L}\bar{g}_{+L}\right)\partial_{6}\left(\sum_{n,q}\psi_{-R}f_{+R}g_{+R}\right)\\
			-\left(\sum_{m,p}\bar{\psi}_{-R}\bar{f}_{-R}\bar{g}_{-R}\right)\partial_{6}\left(\sum_{n,q}\psi_{+L}f_{-L}g_{-L}\right)\\
			-\left(\sum_{m,p}\bar{\psi}_{-L}\bar{f}_{-L}\bar{g}_{-L}\right)\partial_{6}\left(\sum_{n,q}\psi_{+R}f_{-R}g_{-R}\right)\bigg]\\
		-e^{-4\sigma}r_{c}R^{2}\sin \theta M_{7}\epsilon(\phi) \times\\
		\bigg[\left(\sum_{m,p}\bar{\psi}_{+R}\bar{f}_{+R}\bar{g}_{+R}\right)\left(\sum_{n,q}\psi_{-L}f_{-L}g_{-L}\right)\\
			+\left(\sum_{m,p}\bar{\psi}_{+L}\bar{f}_{+L}\bar{g}_{+L}\right)\left(\sum_{n,q}\psi_{-R}f_{-R}g_{-R}\right)\\
			+\left(\sum_{m,p}\bar{\psi}_{-R}\bar{f}_{-R}\bar{g}_{-R}\right)\left(\sum_{n,q}\psi_{+L}f_{+L}g_{+L}\right)\\
			+\left(\sum_{m,p}\bar{\psi}_{-L}\bar{f}_{-L}\bar{g}_{-L}\right)\left(\sum_{n,q}\psi_{+R}f_{+R}g_{+R}\right)\bigg]\Bigg)
	\end{multline}
Note that, in this formula, $\partial_{5}$, $\partial_{6}$, and $\cos \theta$ are odd under $\mathbb{Z}_{2}'$ parity, and $\sin \theta$ is even. We can check by similar arguments as before that the same parity choices (eq.~\ref{Parities}) are consistent with this formula.

Comparing this to the RS1 fermion action (eq.~\ref{Expanded RS1 Action}) yields the following normalization condition:
	\[
		\int_{0}^{\pi}\int_{-\pi}^{\pi} R^{2}\sin \theta \bar{g}_{+L}^{p_{1},p_{2}}g_{+L}^{q_{1},q_{2}}\textrm{d}\omega \textrm{d}\theta=\delta_{p_{1}q_{1}}			\delta_{p_{2}q_{2}}
	\]
As before, comparing the RS1 kinetic terms gives us $g_{-L}=g_{+R}$ and $g_{-R}=g_{+L}$. Comparing the RS1 mass term to the $\partial_{5}$ and 
$\partial_{6}$ kinetic terms, we find the following pair of coupled differential equations:
	\begin{align}
		(\partial_{5}+\frac{1}{2}\cot \theta-i\csc \theta \partial_{6})g_{-L}^{p_{1},p_{2}}=&M_{p_{1},p_{2}}g_{-R}^{p_{1},p_{2}}\\
		(\partial_{5}+\frac{1}{2}\cot \theta+i\csc \theta \partial_{6})g_{-R}^{p_{1},p_{2}}=&-M_{p_{1},p_{2}}g_{-L}^{p_{1},p_{2}}
	\end{align}

After applying all these conditions, the 7D action reduces to 5D exactly as before (eq.~\ref{5D Reduced Action}), and the 5D to 4D KK-reduction is identical. However, there is one crucial difference between these two cases. The solution to the above pair of differential equations, which is detailed in Abrikosov \cite{Abrikosov}, can be expressed in terms of Jacobi polynomials. Unlike the differential equations on the torus, these equations cannot have eigenvalue zero. This lack of fermion zero-modes is a special case of a general result for manifolds of positive curvature. To ensure the presence of a massless zero mode, we cannot allow our fermion to propagate in the spherical extra dimensions without the addition of some extra structure.
	
\section{Conclusions \label{sec:conclusion}}

In this paper, we have successfully replicated the results of McDonald \cite{McDonald1} for fermions on the space $\textrm{AdS}_{5}/\mathbb{Z}_{2}\times T_{2}/\mathbb{Z}_{2}'$. This model allows a chiral fermion zero mode, includes a dark matter candidate, and incorporates the RS1 approach to the Planck-weak and fermion mass hierarchy problems. However, our decomposition of the 8-component fermion into two 4-component fermions is different from the method described in McDonald, and has several conceptual benefits.

In addition, we found that the compactification $\textrm{AdS}_{5}/\mathbb{Z}_{2}\times S_{2}/\mathbb{Z}_{2}'$ is not viable phenomenologically unless some additional structure is specified. In particular, we can confine the fermions to `points' of codimension two, as is detailed (for the case of $\mathbb{M}^{4}\times S^{2}$) in \cite{SphereFermionsConfinedtoPoints}. Alternatively, a new $U(1)$ gauge group can be introduced, cancelling the effects of the positive curvature in the 7D action and allowing a fermion zero-mode \cite{SphereExtraGaugeGroup} (again, this paper deals with $\mathbb{M}^{4}\times S^{2}$). Future research is necessary to show that these problems are still resolved in the seven-dimensional case. In addition, models with two flat extra dimensions have a number of desirable features, including an anomaly-cancellation constraint on the number of generations of fermions \cite{FermionGenerations} and an explanation for the long proton lifetime \cite{ProtonStability}. It is possible that these features also extend to the models examined in this paper.

\begin{acknowledgments}
	We would like to thank Casey Douglas for helpful discussions on topology and orbifolding. 
\end{acknowledgments}

\bibliography{7Dfermions}

\end{document}